\documentclass[aps,prl,superscriptaddress,twocolumn,showpacs]{revtex4-1}
\usepackage[table]{xcolor}
\usepackage{pgf}
\usepackage[utf8]{inputenc}
\usepackage{amsmath}
\usepackage{braket}
\usepackage{amssymb}
\usepackage{amsmath}
\usepackage{bbold}
\usepackage{enumitem} 
\usepackage{appendix}
\usepackage{graphicx}
\usepackage{overpic}
\usepackage[pdftex,bookmarks,colorlinks]{hyperref}
\usepackage{hyperref}

\newcommand{\mean}[1]{\left\langle #1\right\rangle}

\begin{document}
\title{Many-body synchronisation in a classical Hamiltonian system}
\author{Reyhaneh Khasseh}
\affiliation{Department of Physics, Institute for Advanced Studies in Basic Sciences (IASBS), Zanjan 45137-66731, Iran}
\affiliation{Abdus Salam ICTP, Strada Costiera 11, I-34151 Trieste, Italy}

\author{Rosario Fazio}
\affiliation{Abdus Salam ICTP, Strada Costiera 11, I-34151 Trieste, Italy}
\affiliation{Dipartimento di Fisica, Universita di Napoli "Federico II", Monte S. Angelo, I-80126 Napoli, Italy}

\author{Stefano Ruffo}
\affiliation{SISSA, Via Bonomea 265, I-34136 and INFN Trieste, Italy}
\affiliation{Istituto dei Sistemi Complessi -- CNR, Via dei Taurini 19, I-00185 Roma, Italy}

\author{Angelo Russomanno}
\affiliation{Abdus Salam ICTP, Strada Costiera 11, I-34151 Trieste, Italy}
\affiliation{NEST, Scuola Normale Superiore \& Istituto Nanoscienze-CNR, I-56126 Pisa, Italy}

\begin{abstract}
We study synchronisation between periodically driven, interacting classical spins undergoing a Hamiltonian dynamics. In the thermodynamic limit there is a transition between a regime where all the spins oscillate synchronously for an infinite time with a period twice as the driving period (synchronized regime) and a regime where the oscillations die after a finite transient (chaotic regime).  We emphasize the peculiarity of our result, having been synchronisation observed so far only in driven-dissipative systems. We discuss how our findings can be interpreted as a period-doubling time crystal and we show that synchronisation can appear both for an overall regular and an overall chaotic dynamics.
\end{abstract}

\maketitle
%\section{Introduction}
%
{Since its discovery by Huygens, the phenomenon of synchronisation~\cite{Pikovsking,Strogazz,Gupta,arenas} has emerged in the most diverse 
contexts. Examples of systems undergoing synchronised motion range from coupled mechanical oscillators  to chemical reactions, from modulated lasers to 
neuronal networks or circadian rhythms in living organisms, just to mention only few of them. The essence of synchronisation can be very simply stated.
Classical non-linear  systems may asymptotically approach self-sustained oscillations, a tiny coupling between those systems can induce their oscillations to be 
locked in phase space.} 

{All  known systems undergoing synchronised dynamics are driven and dissipative. It is therefore natural to ask if synchronisation can 
occur in a Hamiltonian classical system. This is the problem we will address in this work. As we will see below, this question, besides having a direct 
impact on our understanding of dynamical systems, has important connections to chaos and the the foundations of statistical mechanics.} 

In the case of a finite number of coupled classical Hamiltonian systems whose dynamics is generically chaotic, synchronisation can be ruled out.
For more than two degrees of freedom, even small integrability breaking leads eventually to instability of the motion and chaos. In the many-body case this fact leads to thermalisation (at infinite temperature 
in the driven case)~\cite{Berry,Lichtenberg,Arnold_Avez}. This picture can drastically change if an {\it infinite} number of coupled time-dependent 
classical Hamiltonian systems is considered. In this work we will show that synchronisation is possible in this case. \textcolor{black}{To the best of our knowledge, synchronisation in Hamiltonian systems has not been considered before.}

%\textcolor{black}{The seek for many-body systems exhibiting ergodic (or the lack of it) behaviour goes back to the foundations of statistical  physics. Recently in the 
%quantum realm the study of integrable systems has intensified both theoretically and experimentally and new important examples of ergodicity breaking, as 
%many-body localisation, have been discovered. .} 
{This result has non-trivial connections to the foundations of statistical mechanics. Usually, in the thermodynamic limit, any integrability breaking of short-range 
interacting classical Hamiltonian systems leads to an essentially fully chaotic behaviour and hence to thermalisation~\cite{Pettini,Konishi}. Nevertheless,
this is not the whole story~\cite{Lichtenberg} and there may be important cases where this scenario does not apply. In many-body quantum systems, ergodicity can be broken in an extended region of coupling parameters due to interference effects; relevant examples are many-body localisation 
(see~\cite{MBL_review} for a review) and many-body dynamical localisation~\cite{Rozenbaum,Rylands,Michele}. The challenge is to get a similar stabilisation 
in classical Hamiltonian systems, where the quantum interference provides no help. Here we provide an example in a driven context.}

{ In this letter we consider a system of coupled classical spins undergoing a periodic pulsed driving. 
The key point of our analysis will be considering long-range interacting Hamiltonian systems. Here the dynamics can be equivalent to one single collective degree of 
freedom weakly coupled to other modes and the dynamics  can be regular~\cite{Tsuchiya,Campa:book} in the thermodynamic limit~\cite{footnote}.  
The effect of periodic kicking on the regular/chaotic dynamics of classical Hamiltonian systems has been already widely investigated, see 
e.g.~\cite{Konishi,Oven,Lichtenberg,chiri_vov,Ata_ema,Haake}. Here we make a step forward and analyse the synchronisation behaviour.}

\begin{figure*}
\hspace{-0.5cm}\begin{overpic}[width=110mm]{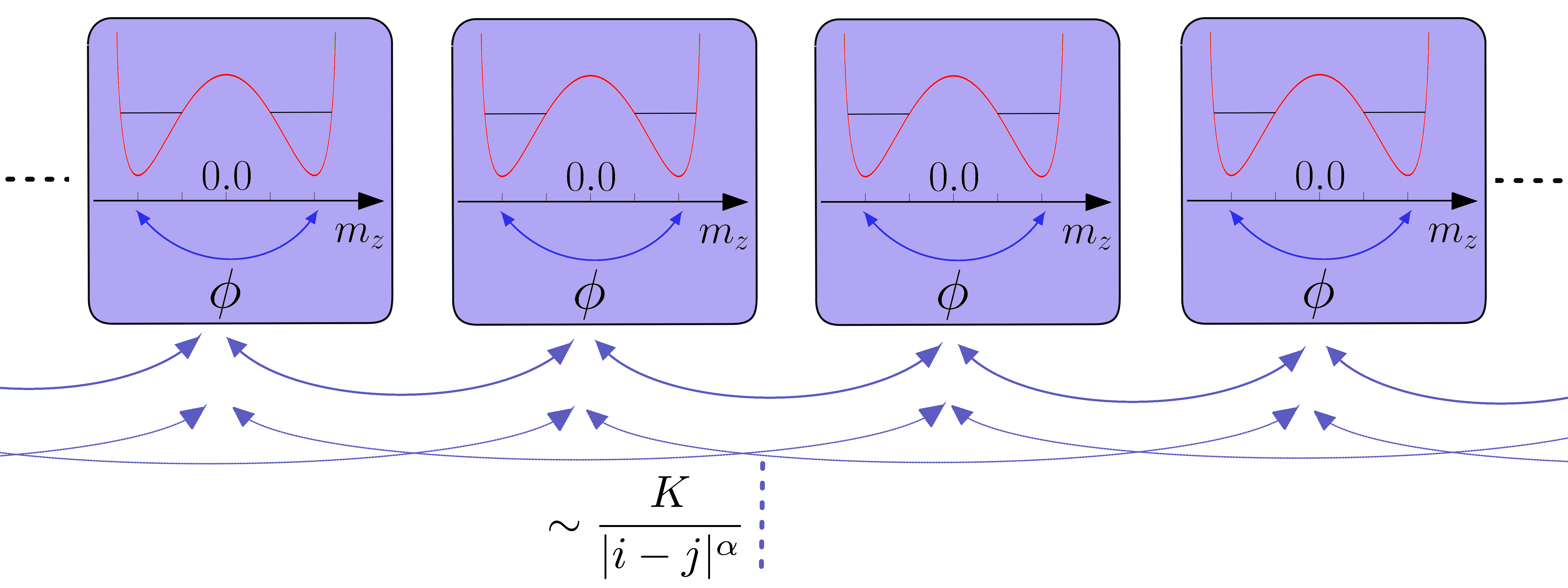}\put(0,40){(a)}\end{overpic}\;
\begin{overpic}[width=70mm]{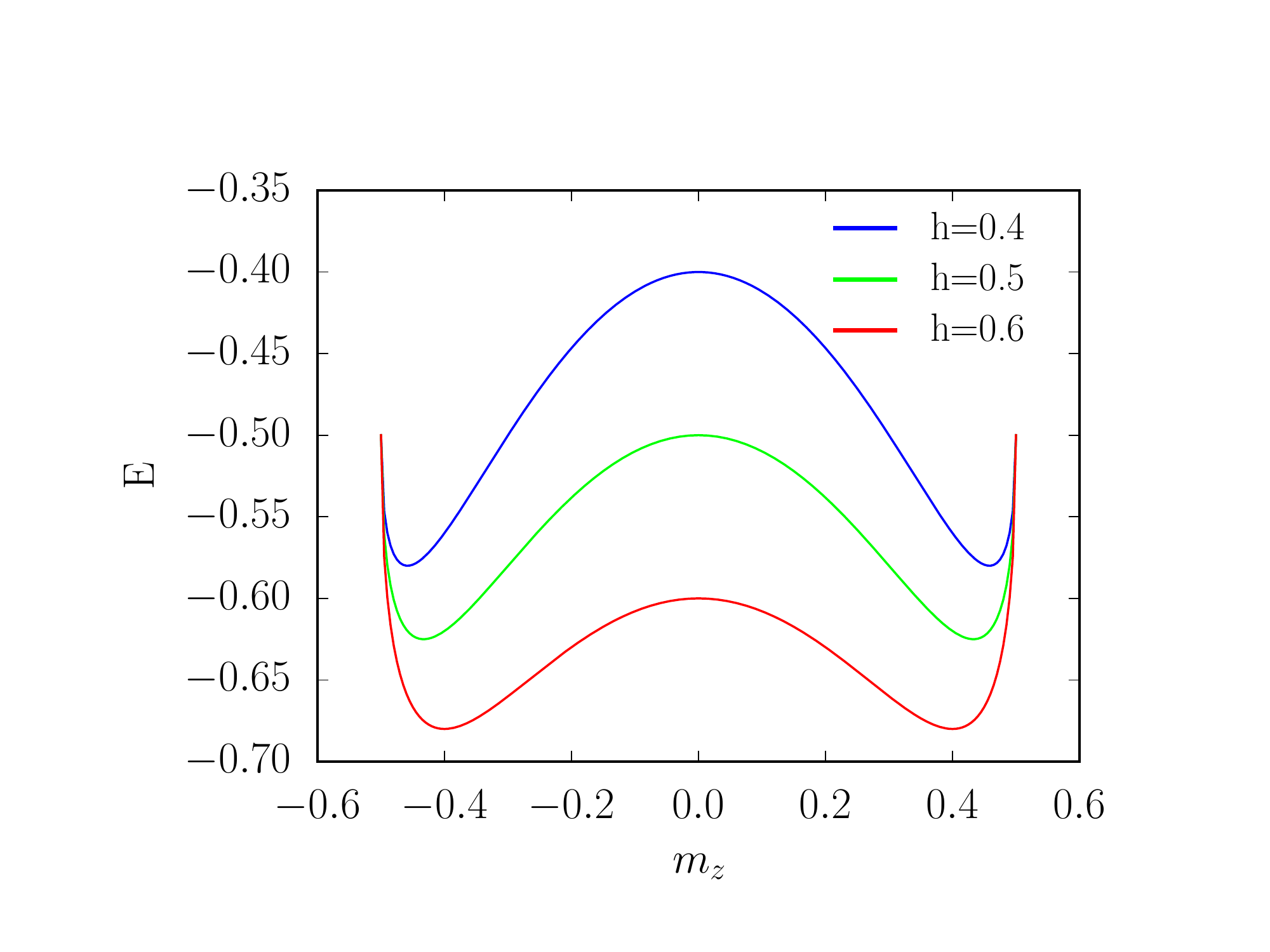}\put(20,75){(b)}\end{overpic}
\caption{(a) Long-range coupled period-doubling classical spin systems. The blue boxes symbolise the single oscillators: Without interactions 
they are entrained with the single-particle kicking of angle $\phi$, swapping the two symmetry sectors at each kick. The long arrows indicate the interacting 
part of the kicking which decays as a power law in the distance, with exponent $\alpha$. If $\alpha$ is small enough there is still synchronisation in the 
thermodynamic limit and appears as period-doubling oscillations in the average $z$-magnetisation. (b) Symmetry-breaking in the phase space for a single 
oscillator (the plots are for $m^y=0$). Around each of the two degenerate minima and for energies smaller than the broken-symmetry edge $E^*$ (the 
central maximum at $m^z=0$) there are trajectories which break the $\mathbb{Z}_2$ symmetry.  }
\label{fig:oscill0}
\end{figure*}

{If uncoupled, the  dynamics of the spins is regular and they show entrainment with the external driving: The magnetisation oscillates with a period double 
of that of the driving field. Once the spins are coupled  through the  driving (see Fig.~\ref{fig:oscill0}), they show synchronised period-doubling oscillations 
for a time which scales with the system size and tends to infinity in the thermodynamic limit.  Therefore synchronisation is an emergent phenomenon, occurring only in 
the thermodynamic limit of an infinite interacting system, much like a  spontaneous symmetry breaking (as the one occurring in the Kuramoto model~\cite{kuramoto,kuramoto1}). 
We remark that the spins are both entrained with the external driving and synchronised with each other. }

Being a form of spatio-temporal order in 
the thermodynamic limit, robust in a full region of the parameter space and for many initial conditions, occurring as a period 
doubling with respect to the driving, synchronization can be interpreted as a spontaneous breaking of the discrete time-translation symmetry (from the symmetry 
group $\mathbb{Z}$ to $2\mathbb{Z}$). %
Indeed, we can see this dynamics as a classical Hamiltonian period doubling inspired by Floquet quantum time crystals (see~\cite{Else,Vedika}). 
\textcolor{black}{Our result is an example of spontaneous time-translation symmetry breaking in a classical Hamiltonian
system.} Until now, the only known examples of classical time crystal are driven-dissipative systems~\cite{Chetan,Gambetta}. 
We remark that other forms of synchronisation could be possible where the non-trivial response of the system has the same period of the driving: 
in this case there would be synchronisation without time-translation symmetry breaking. 

The Hamiltonian governing the $N$ classical spins  $\vec{m}_i$ is given by  ${\cal H}(t)=\sum_{j=1}^N{\cal H}^{(0)}(t) + {\cal V}(t)$.
The non-interacting part has the form
\begin{equation}
	{\cal H}^{(0)}(t) =  \sum_{j=1}^N \left [-2J(m_{j}^z)^2-2h_jm_{j}^x+\phi\,\delta_\tau(t)\, m_{j}^x\right]
\end{equation} 
and the kicked long-range interaction term is
\begin{equation}\label{eq:kick-Hamiltonian-lr}
	{\cal V} (t)=-K\delta_\tau(t) \sum_{i,\,j\neq i}\frac{1}{D_{i,j}^{(\alpha)}}~m_{i}^xm_{j}^x\,,
\end{equation}
where, as in~\cite{chiri_vov}, we define $\delta_\tau(t) \equiv \sum_{n}\delta(t-n\tau)$ to characterise the periodic kicks of period $\tau$ and  $J$, $h_j$, $\phi$ and $K$ are 
tunable parameters. Throughout all the paper we consider periodic boundary 
conditions, we have defined $D^{(\alpha)}_{i,j}\equiv \kappa (\alpha) [\min\left\{|i-j|,N-|i-j|\right\}]^{\alpha}$ in order to implement  them with the same 
prescription of~\cite{gorshkov,mori} ($D^{(\alpha)}_{i,j} \sim \kappa(\alpha)|i-j|^{\alpha}$ when $|i-j| \ll N$). The quantity $\kappa(\alpha)$ is needed in order to make the interaction 
part of the Hamiltonian extensive~\cite{Ekkekkac}: $\kappa(\alpha)\equiv\ N^{1-\alpha}$  if  $\alpha<1$ and unit for $\alpha>1$, in the marginal case $\alpha =1$  equals $\log N $.

The dynamics of this Hamiltonian is  obtained using the Poisson-bracket rules of the classical-spin components  $\left\{m_i^\mu,\,m_j^\nu\right\}=
\epsilon^{\mu\,\nu\,\rho}\delta_{i\,j}m_j^\rho$ where the Greek indices can take values in $x,y,z$, the Latin ones in $1,\ldots,N$ and 
$\epsilon^{\mu\,\nu\,\rho}$  is the Ricci fully antisymmetric tensor. With these Poisson-bracket rules it is easy to write down the Hamilton equations
 $\dot{m}_j^\nu(t)=-\left\{{m}_j^\nu(t),\mathcal{ H}(t)\right\}\,$. Between two kicks  they are a set of $N$ decoupled systems of $3$ differential equations.
 Across a kick they can be explicitly integrated and they give rise for each $j$ to a rotation around the $x$ axis with angle 
 depending on  the values of the $\{m_{l}^x\}$.
% $\theta_j(\{m_l^x\})$ depends on 
% the values of the $\{m_{l}^x\}$ as  
%\begin{equation}
%  	\theta_j(\{m_j^x\})\equiv \phi_j-\frac{2}{N(\alpha)}\sum_{l\neq j}\frac{\lambda}{|l|^\alpha}~m_{j+l}^x\,.
%\end{equation}

Let us start from the case with no interactions ($K \equiv 0$). In this case there is a range of parameters where each classical 
spin can show a period-doubling response to the driving~\cite{FTC_Russomanno}. When $h_j<J$, the Hamiltonian shows a $\mathbb{Z}_{2}$ symmetry breaking. The 
Hamiltonian is indeed symmetric under the $\pi$-rotation around the $x$ axis ($m_l^{y,z}\rightarrow -m_l^{y,z}$, $m_l^x\rightarrow m_l^x$ $\forall$ $l$) but the trajectories with energy 
smaller than a broken-symmetry edge~\cite{fabri} break this symmetry. These trajectories are doubly degenerate and appear in pairs transformed into each other by the symmetry operation
(see Fig.~\ref{fig:oscill0}(b)). The system shows period doubling if it is prepared in a symmetry-breaking trajectory and the kicking with $K\equiv 0$ is used to 
swap between this trajectory and its symmetric partner.
The kick  produces a rotation of angle $\phi$ around the $x$ axis. By considering $\phi\equiv\pi$ there are period-doubling oscillations of the $z$-magnetisations $m_{j}^z$ 
(perfect swapping of the symmetric trajectories). These oscillations are stable if $\phi$ is made slightly different from $\pi$, there being a continuum of symmetry breaking 
trajectories (see Ref.~\cite{FTC_Russomanno}). 

The analysis of the interacting dynamics  $K \ne 0$  is crucial to understand when the period doubling is stable in the thermodynamic limit.
We will characterise the interacting dynamics by analysing the average magnetisation along the $z-$axis, $m^z(t)$ [see Eq.~\eqref{eq:kick-Hamiltonian-lr}]. For any finite size we see period-doubling oscillations of $m^z(t)$.
These oscillations mark the synchronisation of the oscillators and are discrete rotations in time analogous to the continuous ones of the Kuramoto order parameter~\cite{kuramoto,kuramoto1}.  The period-doubling oscillations die out after a transient; %to the the $T=\infty$ thermal value after some time $t_d$. (The $T=\infty$ thermal value is $m^z=0$ and is computed in the canonical ensemble for the Hamiltonian without a kicking.) As we will see below, there is a region in the parameter 
%space where $t_d$ increases as a power law with the system size and there is perfect synchronisation among the spins in the infinite-size limit. %We study also the interplay of 
%synchronisation with chaos, considering the behaviour of the Largest Lyapunov Exponent (LLE). We find a region in the parameter space where 
%synchronisation in the thermodynamic limit can emerge even if the LLE still witnesses a chaotic dynamics. In this regime an essentially regular and an essentially 
%chaotic region of phase space coexist, and synchronisation appears for initial conditions chosen in a specific range of parameters. 
%
in order to see how this transient scales with the system size, we define the order parameter for period doubling
\begin{equation}\label{eq:kick-Hamiltonian-lr}
{O}(n\tau)\equiv(-1)^n m^{z}(n\tau)=\frac{(-1)^n}{N}\sum_{j=1}^Nm_j^z(n\tau)~,
\end{equation}  
where $m^z(t)$ is the average $z$ magnetisation. ${O}(n\tau)$ remains non-vanishing keeping its sign until there are period-doubling oscillations of the spins. For any 
finite size of the system, we numerically see that this quantity vanishes after a transient, reaching in this way the thermal $T=\infty$ value $O_{T=\infty}=m^z_{T=\infty}=0$. (The $T=\infty$ thermal values are computed in the microcanonical ensemble for the Hamiltonian without kicking.) To study the scaling of the 
transient, we quantify its duration as
%
%\begin{equation} \label{t_d:eqn}
$
  t_{d}/\tau= \sum_{n=1}^{t^*/\tau} n~{O}(n\tau)/ \sum_{n=1}^{t^*/\tau}{O}(n\tau)~.
$
%\end{equation}
%
Here $t^*/\tau$ is the first value of $n$ where ${O}(n\tau)$ vanishes. %(we prefer do not use directly $n_{max}$ in order to reduce the fluctuations). 
In order to have persistent synchronized period-doubling oscillations in the thermodynamic limit, $t_d$ must diverge with the system size $N$. 

%Because we are interested in the dynamics of the $z$-component of the magnetisation, w
We initialise the system in a state where the order parameter, $O(0)$, is positive. A uniform initial state is a very singular case: it is easy to show that for a 
uniform Hamiltonian the dynamics is equivalent to a single spin. The synchronisation is trivial and corresponds to the entrainment of the single oscillator. A 
nontrivial situation arises in the case of a random initial state ($m_j^z(0)=\sqrt{1-\epsilon_j^2}$, $m_j^x(0)=\epsilon_j\cos\varphi_j$, $m_j^y(0)=\epsilon_j\sin\varphi_j$ 
with $\epsilon_j$ a random variable uniformly distributed in the interval $[0,\epsilon]$ and $\varphi_j$  uniformly distributed in $[0,2\pi]$). 
We can also include disorder in the Hamiltonian by taking the % the angles $\phi_j$ can be uniformly equal to some 
$h_j$ random and uniformly distributed in the interval $[h-\Delta h,h+\Delta h]$. In these random cases we average our results over $N_{\rm rand}$ randomness realizations and evaluate the errorbars of any randomness-dependent quantity $\mathcal{S}$ as \textcolor{black}{$\operatorname{std}(\mathcal{S})/\sqrt{N_{\rm rand}}$} where $\operatorname{std}(\mathcal{S})$ is the standard deviation of $\mathcal{S}$ over randomness realizations. %The same can happen for the local fields which can be uniformly equal to some value $h$ or uniformly distributed in the interval $[h_0-\Delta h,h_0+\Delta h]$. 

\begin{figure}
%\centering
\begin{tabular}{c}
\begin{overpic}[width=73mm]{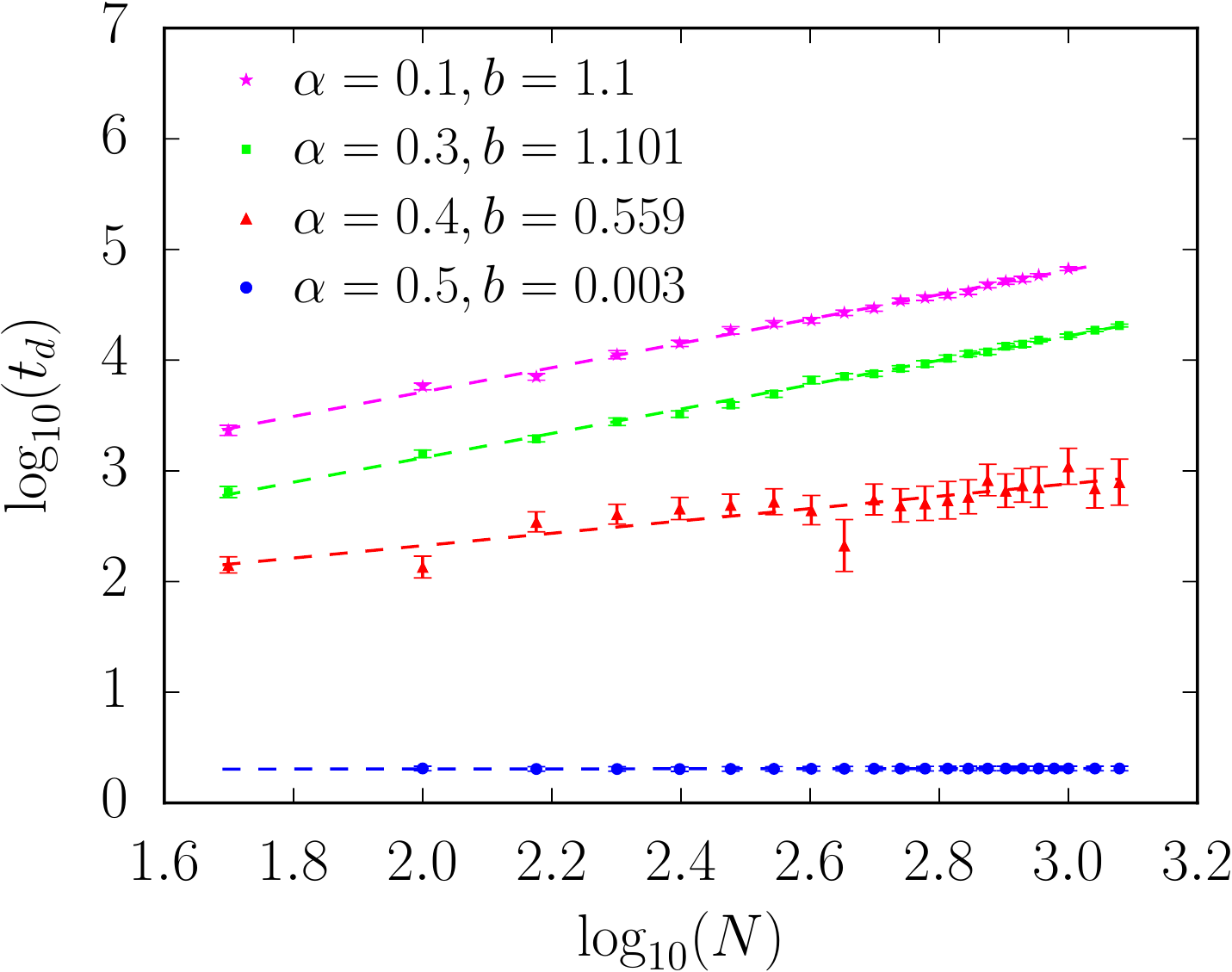}\put(87,70){(a)}\end{overpic}\\
\begin{overpic}[width=75mm]{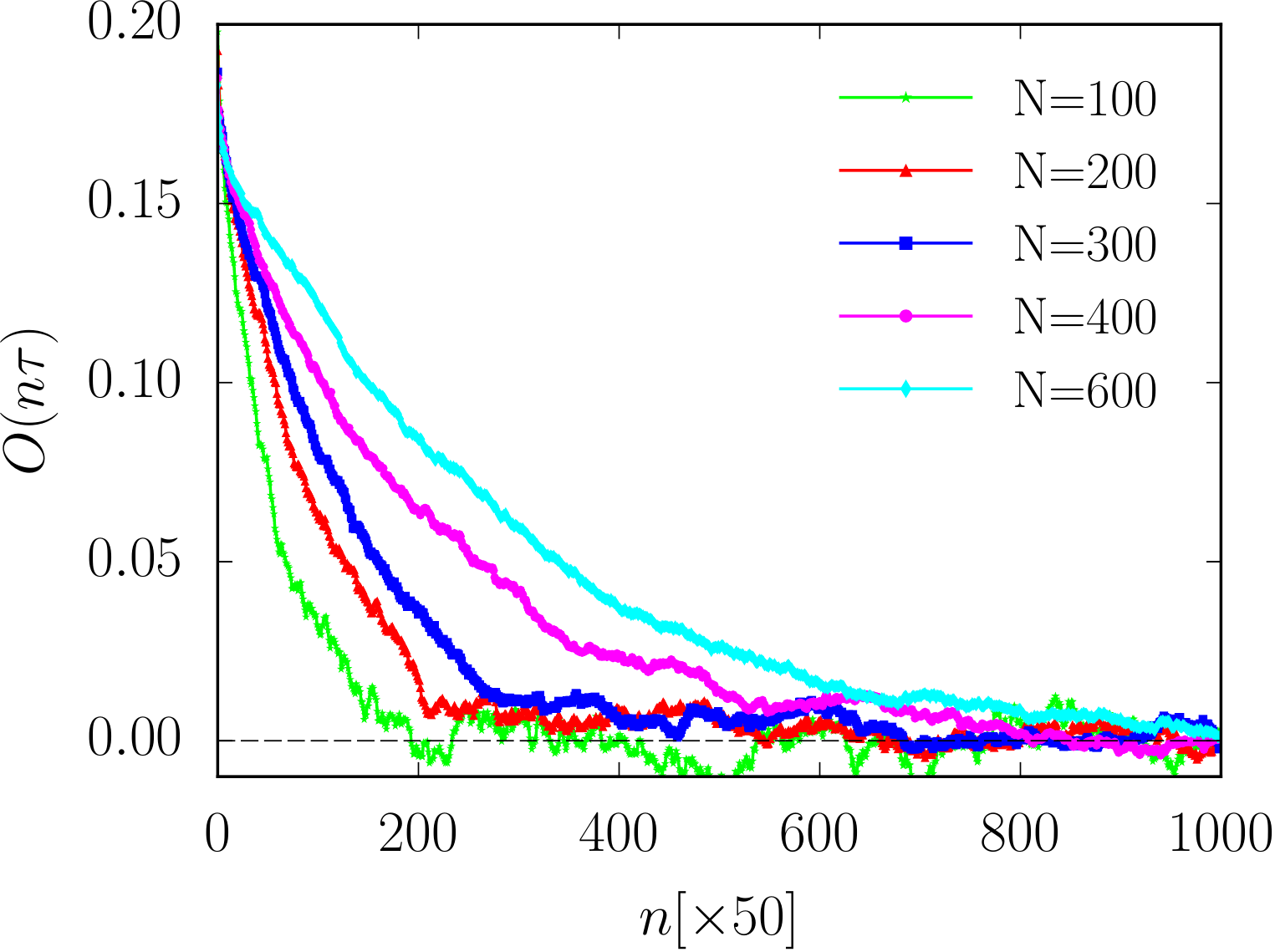}\put(23,65){(b)}\end{overpic}\\
\begin{overpic}[width=75mm]{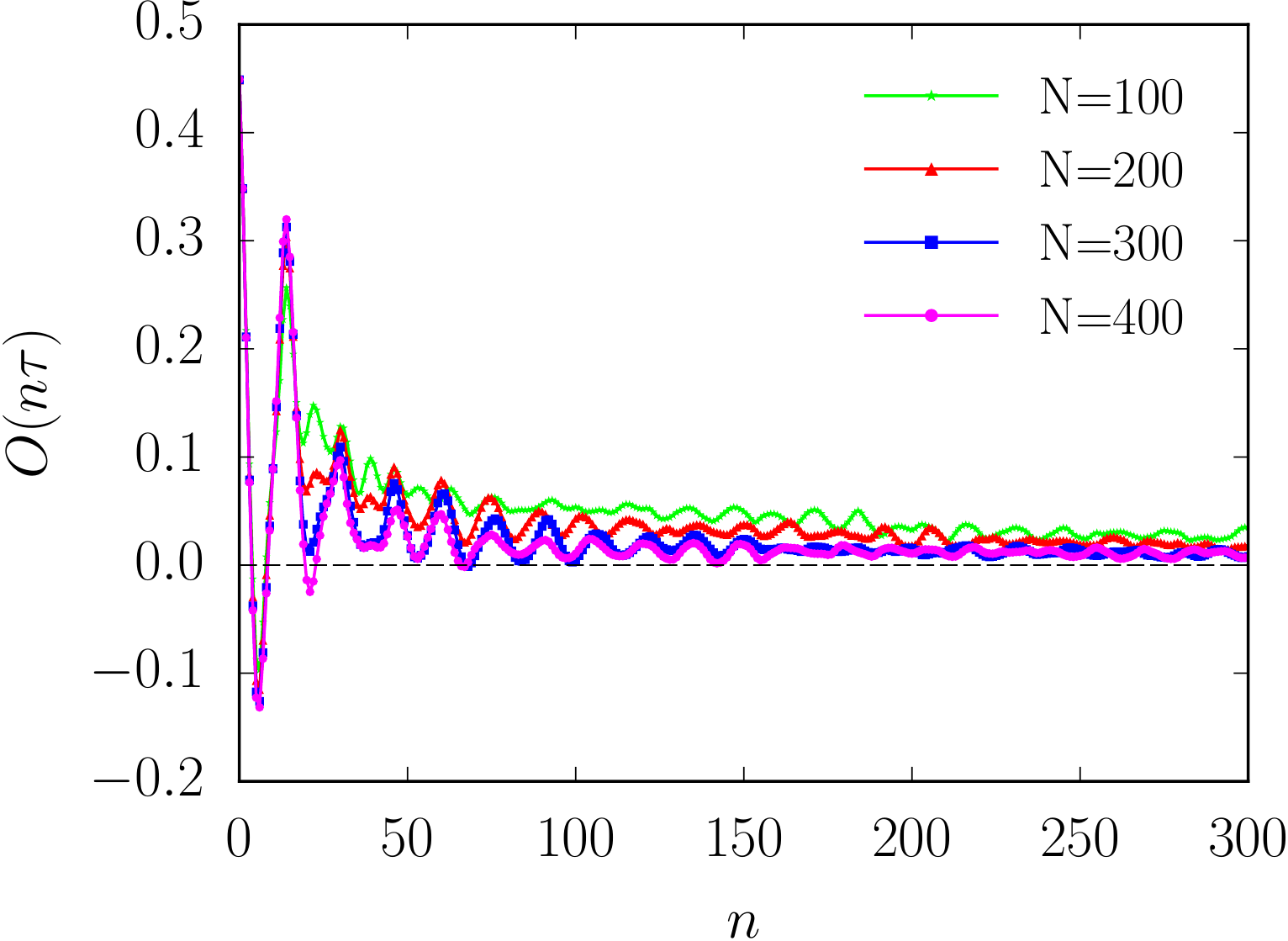}\put(23,62){(c)}\end{overpic}
\end{tabular}
\caption{ (a) Scaling of $t_d$ with $N$. Notice two possible regimes, one in which there is the power-law scaling $t_d\sim N^b$, and another where 
there is no scaling.  (b) Some examples of evolution of the order parameter for different values of $N$ and parameters inside the synchronised region 
($\alpha=0.3$). (c) The same for parameters in the thermalising regime ($\alpha=0.5$). Numerical parameters: $h=0.32$,  $\Delta h = 0$, $\phi=0.99\pi$, 
$K=0.3$, $\tau=0.6$, $\epsilon=0.05$, {$N_{\rm rand}=28$}. }
\label{fig1:tdscaling}
\end{figure}

\begin{figure}
%\centering
\begin{overpic}[width=80mm]{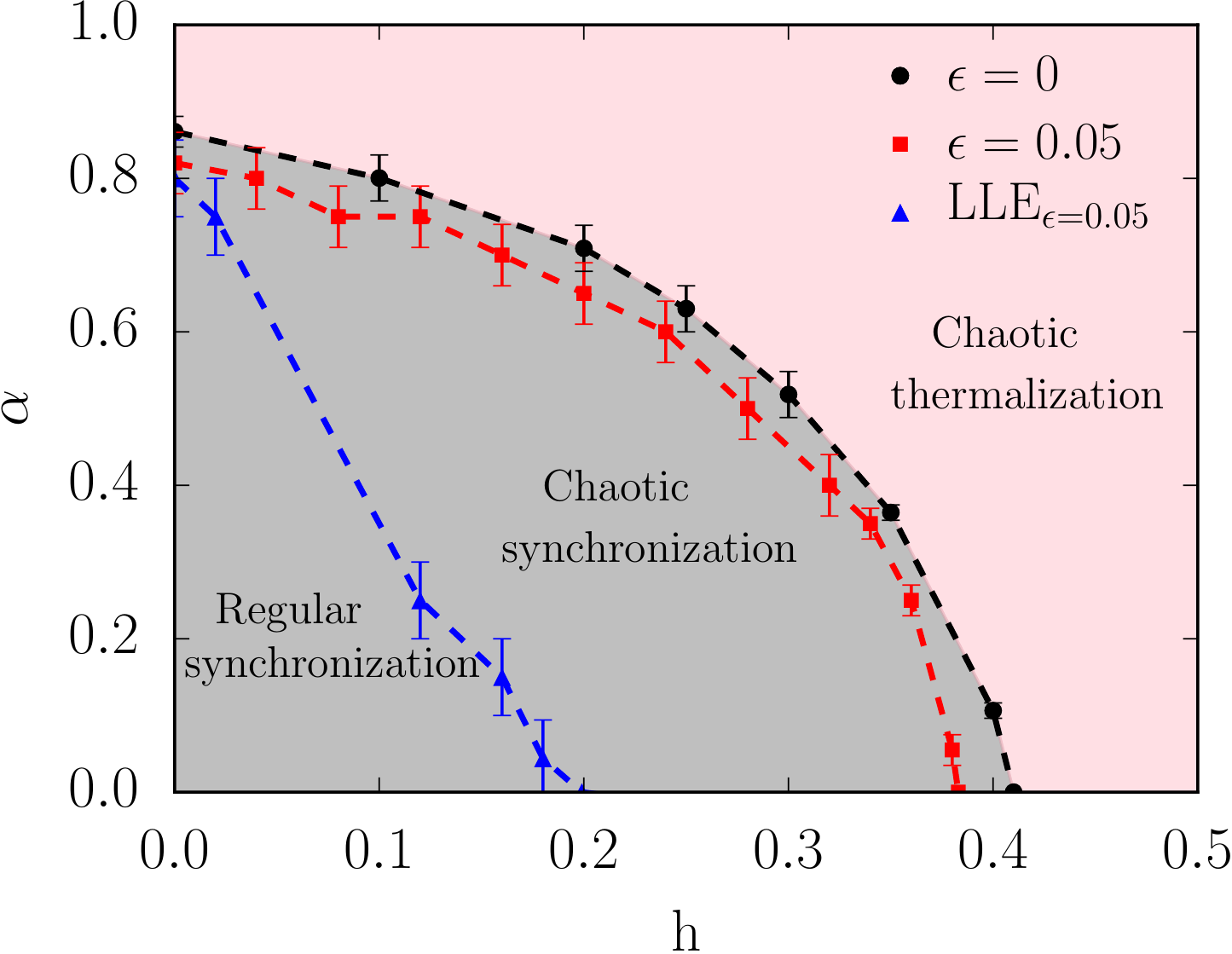}\put(20,70){}\end{overpic}
%\begin{overpic}[width=80mm]{pictures/Fig1b-crop}\put(20,70){(b)}\end{overpic}
\caption{Regions in the parameter space. The red and black curves separate synchronisation from thermalisation at infinite size for different $\epsilon$. The blue curve separates regular behaviour (${\rm LLE}_{\epsilon=0.05}\to0$ in the thermodynamic limit) from chaotic one. Notice the existence of an intermediate chaotic but non thermalising region where ${\rm LLE}_{\epsilon=0.05}>0$ and there is also synchronisation. (Numerical parameters: $K=0.3$, $\tau=0.6$, $\phi\equiv0.99\pi$, $\Delta h = 0$, $N_{\rm rand}=28$.)}% \textcolor{black}{Please Reyhaneh, put the error-bars (the existence of regularity without synchronisation seems to me very strange). Instead of ``DTC'' put ``Regular regime''. Instead of ``Normal'' put ``Ergodic regime''. In the intermediate region put ``Intermediate regime''. Put also a label across the blue line where you write ``synchronisation''. Write ``Random'' in the caption of the red line. } }
%\pgfputat{\pgfxy(5.5,8.9)}{\pgfbox[left,top]{\footnotesize chaotic and}}
\label{fig2:phase_diagram}
\end{figure}

As we vary the parameters of the system we find two regimes. In the synchronised regime the decay time of the order parameter scales as a 
power-law, $t_d\sim N^b$, and there is synchronisation in the thermodynamic limit. In the thermalising regime, on the opposite, there is not such a 
scaling and consequently no synchronisation. Some examples of these two different scalings are shown in Fig.~\ref{fig1:tdscaling}(a). Here we have 
considered the case of a kicking different from the perfect-swapping one (we take $\phi=0.99\pi$ instead of $\phi=\pi$): synchronisation persists also 
for this imperfect kicking, marking thereby the robustness of this phenomenon. {Notice that well inside the synchronized regime the scaling exponent $b$ is very near to 1 and consistent with a linear scaling.}
% but similar features can be found also for values of the phase slightly different than this (we have checked $\phi=0.99\pi$). 
In order to show the markedly different behaviour in the two regimes, in Fig.~\ref{fig1:tdscaling}(b) we provide some examples of evolution of the order 
parameter for different sizes in a case where there is synchronisation and in Fig.~\ref{fig1:tdscaling}(c) we do the same for a case where the system 
thermalises.
%...............................................................................%
%'''''''''''''''''''''''''''''''''''''''''''''''''''''''''''''''''''''''''''''''%
Using the scaling properties of $t_d$, we can clearly distinguish in the thermodynamic limit $N\to\infty$ the synchronised regime from the thermalising 
one and we can map a diagram of the dynamical regimes. %(see SM for technical details). 
We plot this diagram in Fig.~\ref{fig2:phase_diagram} for uniform ($\epsilon=0$, trivial) and random ($\epsilon\neq 0$, non-trivial) 
initial conditions. 
%We see that the value of $\alpha$ where the transition between regularity and chaoticity occurs depends on $h$ and $\lambda$. 
%We have checked that the transition point in $\alpha$ tends to 1 in the limit of $\lambda,h\to 0$, getting thereby in this limit a result equal to the 
%generalized Hamiltonian mean field model~\cite{Celia, Firpo1}.

%...............................................................................%
\begin{figure}
\centering
\begin{tabular}{c}
  \begin{overpic}[width=75mm]{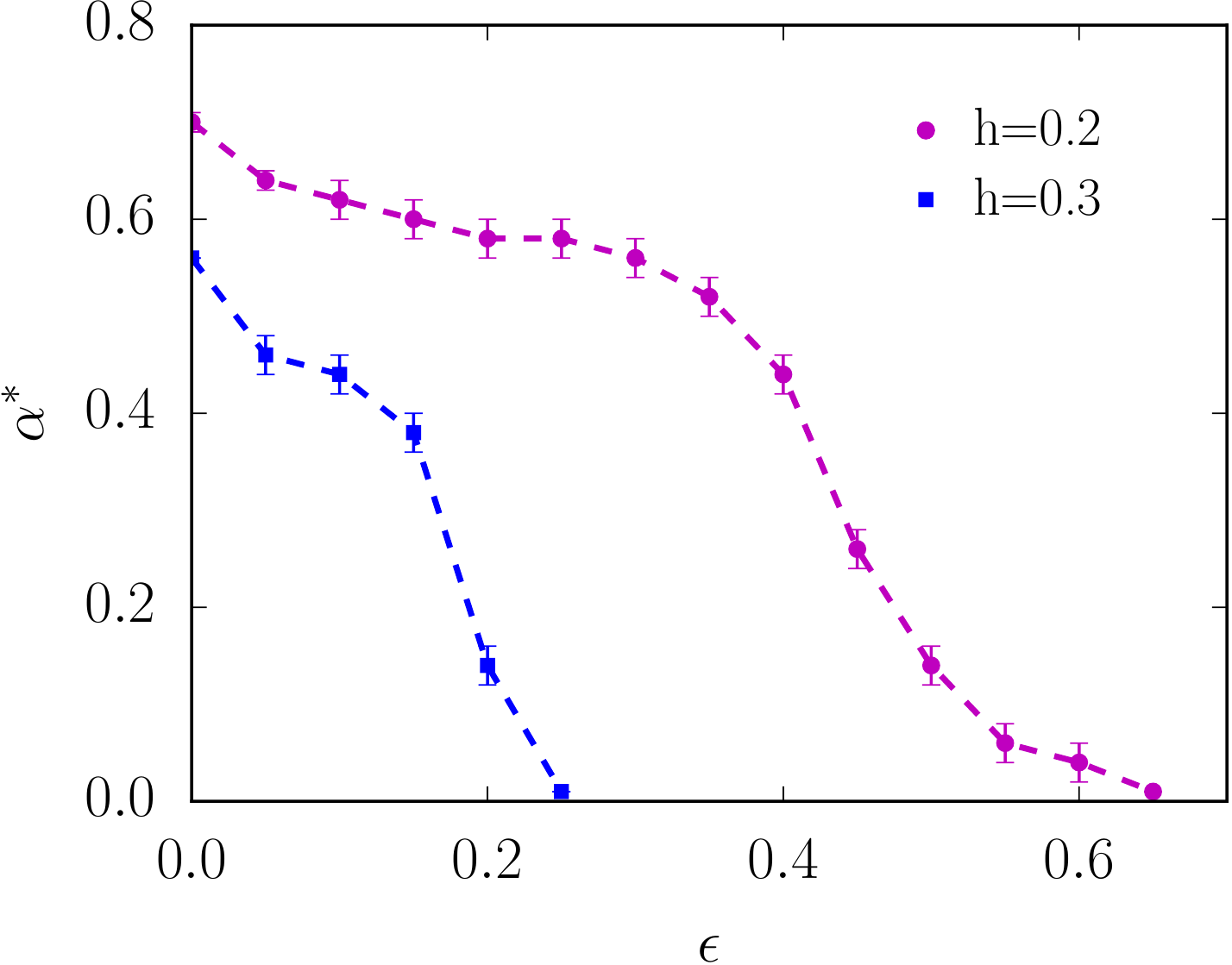}\put(20,70){(a)}\end{overpic}\\
  \begin{overpic}[width=75mm]{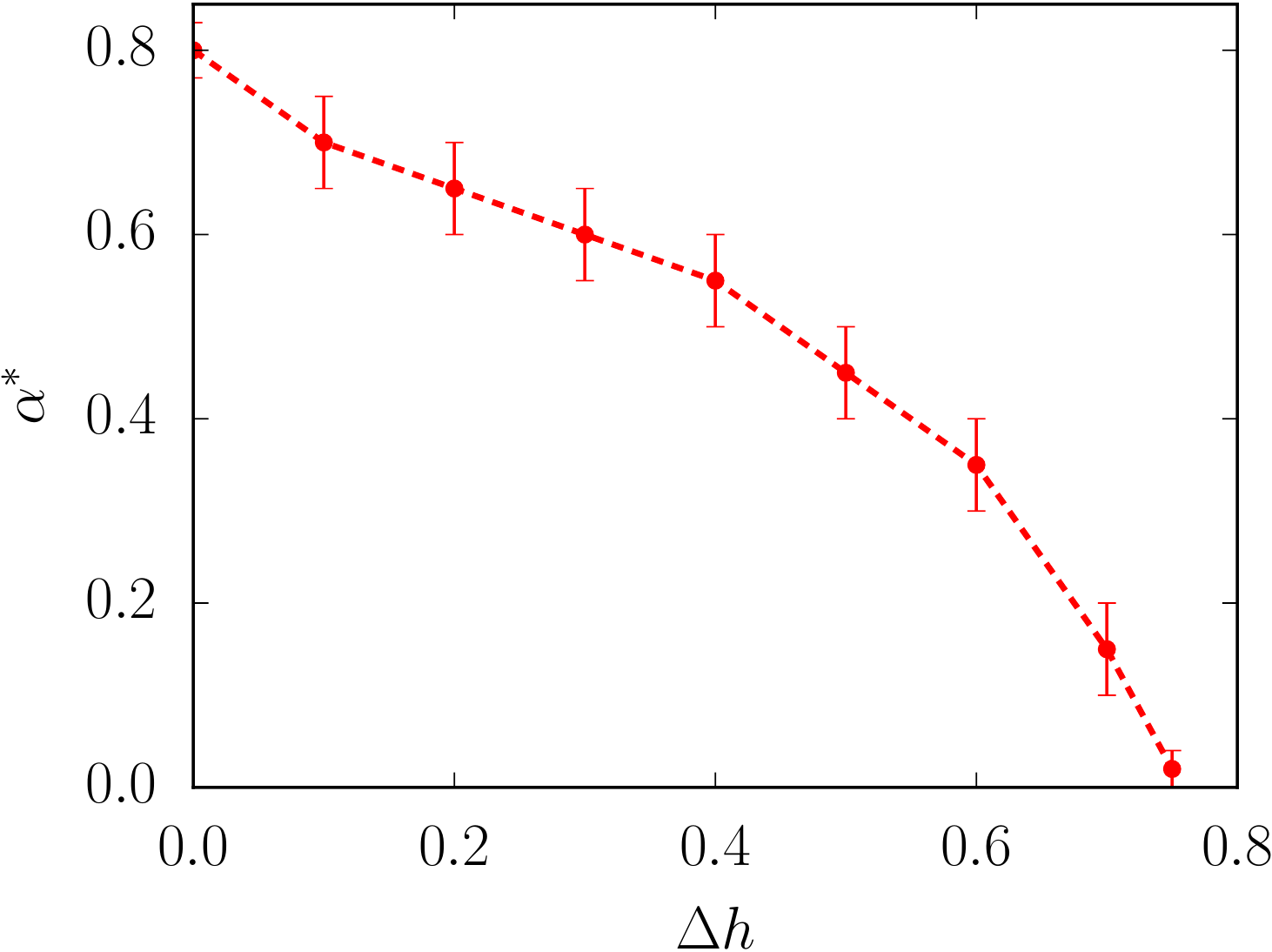}\put(87,70){(b)}\end{overpic}
\end{tabular}
 \caption{Transition point $\alpha^*$ versus the initial-state randomness $\epsilon$ [panel (a)] and the randomness in the field $h_j$ [panel (b)]. Numerical parameters: $K=0.3$, $\tau=0.6$, $\phi=0.99\pi$, {$N_{\rm rand}=20$}; for (a) $\Delta h = 0$, for (b) $h=0.1$.}
%\begin{overpic}[width=80mm]{pictures/Fig1b-crop}\put(20,70){(b)}\end{overpic}\pgfputat{\pgfxy(0.2,6.1)}{\pgfbox[left,top]{\footnotesize *}}
\label{syncoma}
\end{figure}
%...............................................................................% \textcolor{black}{(Reyhaneh, is this correct?)}
 %In Fig.~\ref{slope_alpha:fig} we plot the dependence of $b$ on $\alpha$: we see that there is a point where it abruptly moves from a value larger than 1 to a value smaller than 1. This drop marks the transition point between the synchronized and non-synchronized regime.%The transition is very sharp as can be observed from the dependence of the $t_d$-scaling exponent $b$ on the parameters: there is a sharp discontinuity of $b$ at the phase booundary (see an instance of this in Fig.~\ref{slope_alpha:fig}).

%\begin{figure}
%\begin{overpic}[width=80mm]{pictures/Slope-alpha-h032-crop.pdf}\put(87,70){}\end{overpic}
%\caption{The $t_d$-scaling exponent $b$ versus $\alpha$, notice the abrupt decay at the transition point from synchronisation to chaoticity. Numerical parameters: $h=0.32$, $\phi_j\equiv\phi=\pi$, $\tau=0.6$, uniform initialization.} % \textcolor{black}{(Is this correct?) Reyhaneh, PUT THE ERRORBARS. Moreover, do more points in the interesting transition region. In the end take $\alpha\in[0,1.5]$}. }
%\label{slope_alpha:fig}
%\end{figure}
%.................................................................................................................................................%
%
We remark that synchronisation is  robust and survives the randomness in the initial state. To better show this fact, in Fig.~\ref{syncoma}(a)  we plot $\alpha^*$ (the 
critical value separating synchronised from chaotic and ergodic) versus the randomness amplitude $\epsilon$ for different values of $h$. 
synchronisation  is also robust if disorder is added to the model, as it occurs for example in the Kuramoto model~\cite{kuramoto,kuramoto1,Pikovsking}. We 
have checked this, adding disorder to $h_j$. 
%by taking $\phi_j\in[\phi_0-\Delta\phi,\phi_0+\Delta\phi]$ uniformly distributed, and adding disorder to the local field, taking $h_j\in[h_0-\Delta h,h_0+\Delta h]$ uniformly distributed. 
The results are shown in Fig.~\ref{syncoma}-(b) where we plot the value $\alpha^*$ as a function of  the disorder strength  $\Delta h$.

Let us now move to consider the regularity/chaoticity properties of the dynamics. The largest Lyapunov exponent (LLE)  gives a measure of how much nearby trajectories diverge 
exponentially and is thereby a measure of chaos~\cite{Lyapunov_book}. It is defined as 
${\rm LLE} =\lim_{d(0) \to 0}\lim_{t \to \infty}
\frac{1}{t} \ln\frac{d(t)}{d(0)}$ ($d(t)$ is the distance between trajectories at time $t$). We compute the LLE using the orbit separation method (see~\cite{Kolmogorov_entropy,Lyapunov_book}). We consider its average over the random-initial-conditions distribution introduced above: ${\rm LLE}_\epsilon\equiv\mean{\rm LLE}_{\varphi_j\in[0,2\pi],\,\epsilon_j\in[0,\epsilon]}$. In this way we fix the same distribution of the random initial conditions and here we can compare the regularity/chaoticity properties of the dynamics with the synchronisation properties.

For $N$ finite we find that ${\rm LLE}_\epsilon$
is always larger than 0, as expected for a non-linear non-integrable system, but we can notice two different behaviours in the limit $N\to\infty$ (in the numerics we have fixed $\epsilon=0.05)$. 
There is a regime where ${\rm LLE}_\epsilon$ stays finite in the limit $N\to\infty$ %(see examples in Fig.~\ref{fig1:lyapo}(a)) 
and another regime where our numerics suggests that it scales to 0 as a power law when $N\to\infty$: ${\rm LLE}_\epsilon\sim N^{-\gamma_\epsilon}$ with $\gamma_\epsilon>0$ (as it occurs for the full LLE in the Kuramoto model~\cite{PRE}). We show some examples in the Supplementary Material.% (see Fig.~\ref{LLEN:fig}). % (see examples in Fig.~\ref{fig1:lyapo}(a)). 
We can mark the boundary between the two regimes and plot it as a blue curve in Fig.~\ref{fig2:phase_diagram}. We see that the regular 
region of vanishing ${\rm LLE}_\epsilon$ is smaller than the synchronised region. This suggests that there are three regions in the parameter space for the considered $\epsilon$. {\it Regular synchronisation}: There 
is synchronisation and the ${\rm LLE}_\epsilon\to0$ in the thermodynamic limit. In this case the $N\to\infty$-dynamics is essentially regular in the region of phase space corresponding to the considered random initial conditions. {\it Chaotic thermalisation}: Here ${\rm LLE}_\epsilon>0$ and there is no synchronisation. The dynamics here is essentially chaotic. {\it Chaotic synchronisation}: There is chaos in the considered region of phase space (${\rm LLE}_\epsilon>0$) but forms of order like synchronisation 
can emerge, in analogy with a related phenomenon of a driven-dissipative system~\cite{boris}. We remark that the regularity/chaoticity and synchronisation properties of the dynamics depend on the region of phase space we consider (given by the value of $\epsilon$). We can see this in Fig.~\ref{syncoma}(a) where synchronisation disappears beyond a threshold in $\epsilon$.

In conclusion we have found a form of synchronisation of a set of classical Hamiltonian oscillators which are driven and long-range interacting.  
%In the region of the parameter space where the synchronisation occurs, the time over which the 
synchronisation corresponds to collective period-doubling oscillations lasting for a time which scales as 
a power law with the system size. The synchronisation is robust to randomness in the Hamiltonian and the initial state and is connected to the 
time-crystal phenomena. %We have seen that the LLE scales to 0 as a power law in a parameter region smaller than the one where 
%there is synchronisation, marking the existence of a regime both chaotic and synchronising. 
Perspectives of future research include the  analysis of 
%synchronisation in models with no static symmetries and ratios of response period and driving period larger than 2. One very interesting point is the 
%analysis of 
quantum effects; indeed there are examples of quantum spins with long-range interactions which do not synchronise~\cite{lukin}. It is 
interesting to understand if this phenomenon can be interpreted classically or quantum effects are crucial. {It is also important to consider the role of thermal noise. The situation is very well known for noisy dissipative models with short range interactions~\cite{mukamel1,mukamel2}: Noise generically destroys period $n$-tupling for $n>2$. Noisy dissipative long-range systems have yet to be explored from this perspective. In our specific model we think that thermal noise would spoil synchronisation, 
but this might not be a general feature for long-range systems, especially moving towards the thermodynamic limit.}
\acknowledgements{We acknowledge useful discussions with V.~Latora, S.~Marmi and D.~Mukamel. This work was supported in part by European Union through 
QUIC project (under Grant Agreement 641122). }
%
%............................................................................................................................................................%	
%\appendix
%\section{Calculation of the $T=\infty$ value of the energy} 
%\label{app:thermalisation_energy}
%We want to evaluate the thermal value of $\frac{1}{N}\sum_j\mathcal{H}_{0,\,j}$. Because this object is factorized, we can consider the phase space of a single oscillator.
%To calculate the thermal value we consider the canonical ensemble in which energy of the system per particle can be evaluated as
%
%\begin{equation}
%<H>=\frac{\int_{-1}^{1}dq\int_{0}^{1}dp~\mathcal{H}_0(q,p)e^{-\beta \mathcal{H}_0(q,p)}}{\int_{-1}^{1}dq\int_{0}^{1}dp~e^{-\beta \mathcal{H}_0(q,p)}}~,
%\end{equation} 
%
%\vspace{3mm}

%Where we have used the mapping to canonical coordinates (see for instance~\cite{FTC_Russomanno})
%
%\begin{eqnarray}\label{eq:mag_classical}
%&&m^{x}=~\frac{\sqrt{1-q^2}}{2}\cos (2p),\nonumber \\
%&&m^{y}=~\frac{\sqrt{1-q^2}}{2}\sin (2p),\nonumber \\
%&&m^{z}=\frac{q}{2}~.
%\end{eqnarray}
%
%By using above equations and taking $\beta =0$ we have
%
%\begin{align}
%<\mathcal{H}_0>&=\frac{1}{2\pi}\int_{-1}^{1}\int_{0}^{1}dqdp~\Big(-\frac{1}{2}Jq^2-h\sqrt{1-q^2}\cos (2p)\Big)
%\nonumber\\
%&=-J/6~.
%\end{align} 
%...........................................................................................................................%

%
\clearpage
%\newpage
\clearpage
\setcounter{equation}{0}%
\setcounter{figure}{0}%
\setcounter{table}{0}%
\renewcommand{\thetable}{S\arabic{table}}
\renewcommand{\theequation}{S\arabic{equation}}
\renewcommand{\thefigure}{S\arabic{figure}}

\newcommand\norm[1]{\left\lVert#1\right\rVert}

\begin{center}
  {\Large Supplementary Information}
\end{center}
\section{Scaling of ${\rm LLE}_\epsilon$ with the system size}
\begin{figure}
\centering
\begin{tabular}{c}
  \begin{overpic}[width=75mm]{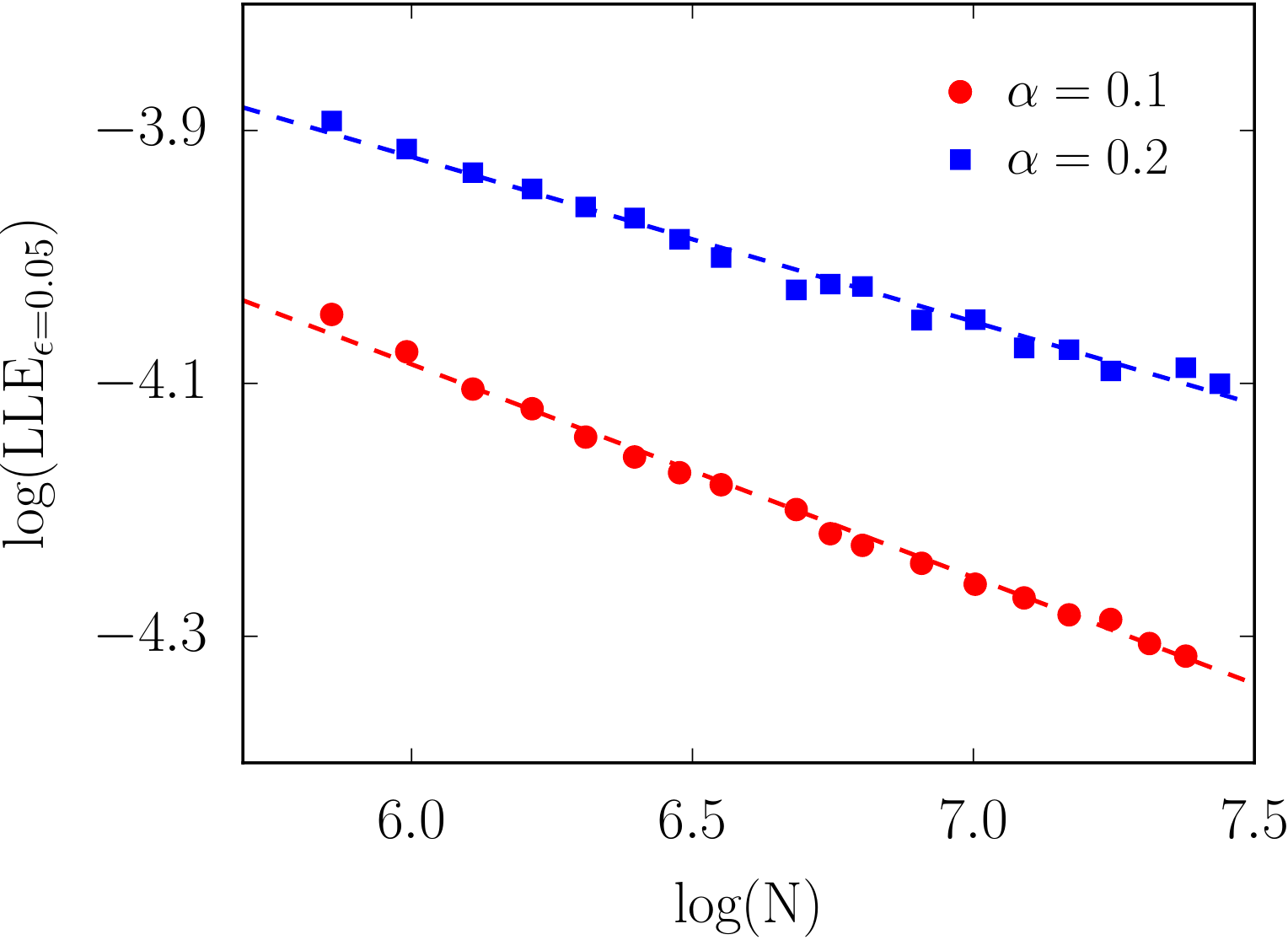}\put(20,67){(a)}\end{overpic}\\
  \begin{overpic}[width=75mm]{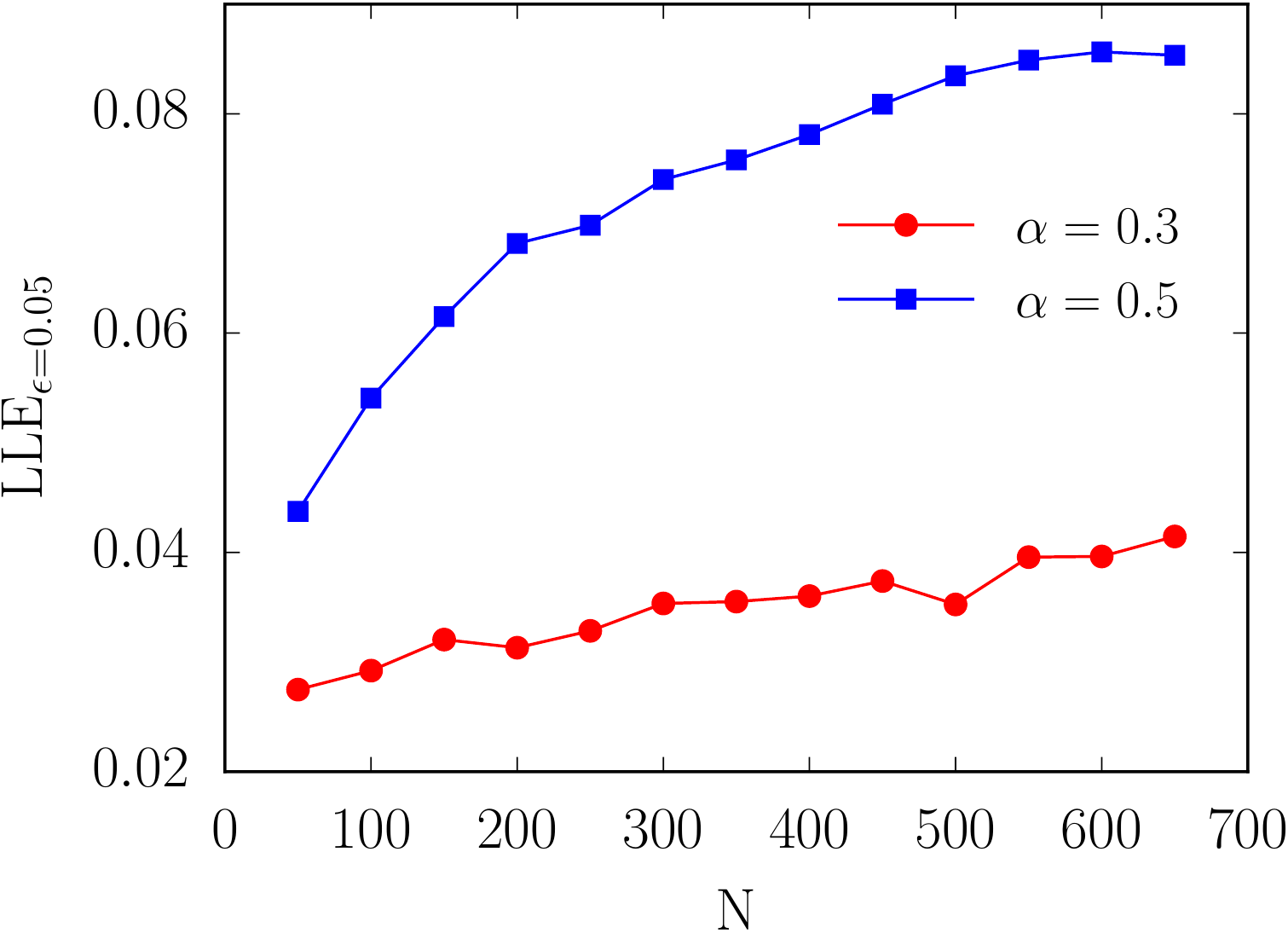}\put(20,67){(b)}\end{overpic}
\end{tabular}
 \caption{${\rm LLE_{\epsilon=0.05}}$ versus $N$ in cases of regular synchronisation (upper panel) and chaotic synchronisation (lower panel). Numerical parameters:  $K=0.3$, $\tau=0.6$, $\phi\equiv0.99\pi$, $\Delta h = 0$, $\epsilon=0.05$, $N_{\rm rand}=28$.}
\label{LLEN:fig}
\end{figure}

\begin{figure}[b]
\centering
   \includegraphics[width=75mm]{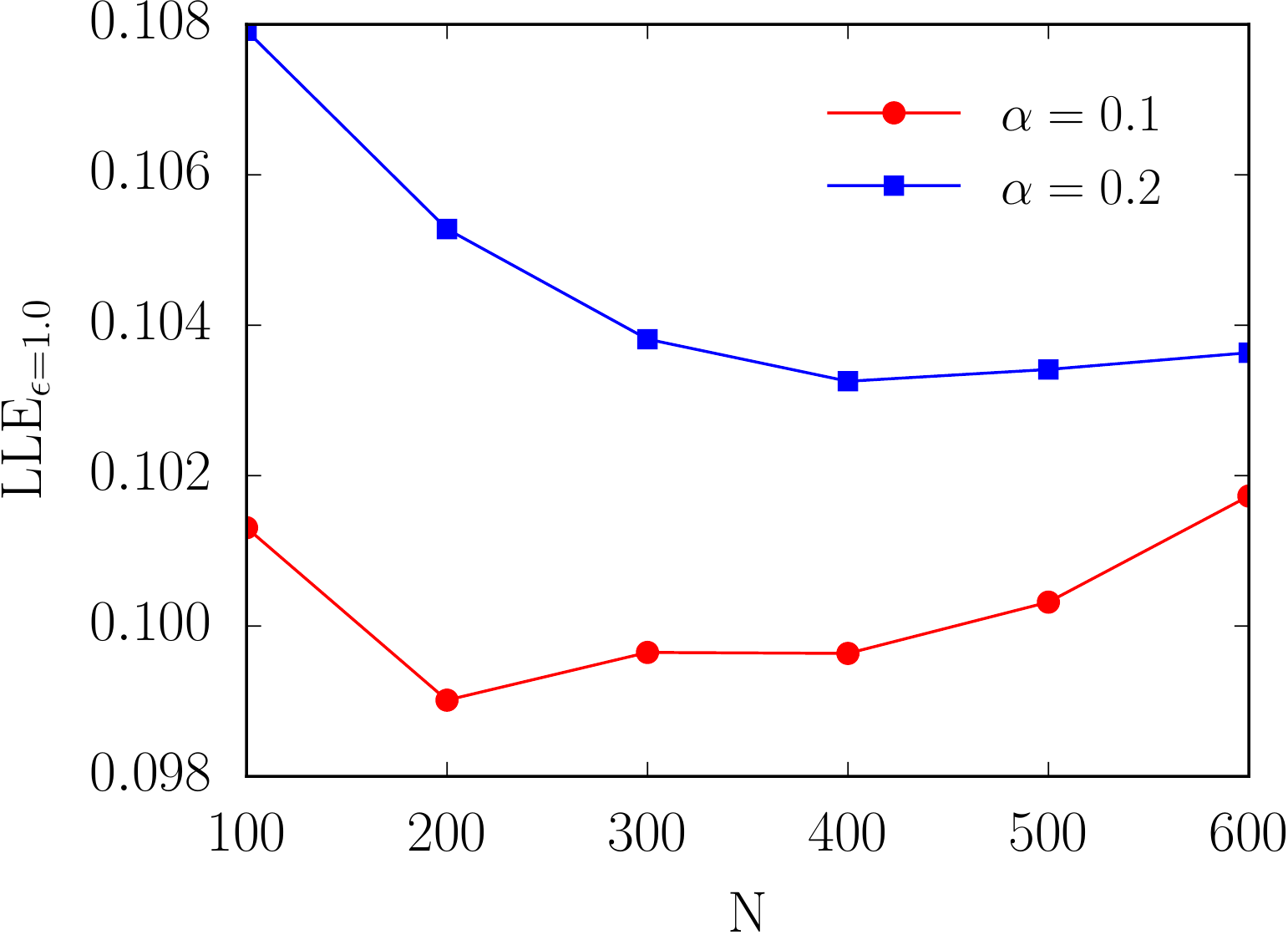}
 \caption{${\rm LLE_{\epsilon=1}}$ versus $N$ in cases where there is regular synchronisation for $\epsilon=0.05$. Numerical parameters:  $K=0.3$, $\tau=0.6$, $\phi\equiv0.99\pi$, $\Delta h = 0$, $\epsilon=1$, $N_{\rm rand}=28$.}
\label{LLEN1:fig}
\end{figure}

We provide here some examples of scaling of ${\rm LLE}_\epsilon$ with $N$. First, we consider cases in the regular-synchronisation region. Here, our numerics suggests a power-law scaling with the system size: ${\rm LLE}_\epsilon\sim N^{-\gamma_\epsilon}$ with $\gamma_\epsilon>0$ [see Fig.~\ref{LLEN:fig}(a) for $\epsilon=0.05$]. Unfortunately, we can reach too small system sizes and we cannot do a statement sharper than ``suggest''. We remark that the existence of this power-law decay strongly depends on the choice of $\epsilon$. Taking a larger $\epsilon$ ($\epsilon=1$ in Fig.~\ref{LLEN1:fig}) there is no more decay. The point is that $\epsilon$ marks the size of the region of phase space where we are probing the regularity/chaoticity behaviour. With small $\epsilon$ we restrict to a regular region of phase space; with larger $\epsilon$ we embrace also the chaotic part of the phase space.

On the opposite, in the chaotic-synchronisation region, ${\rm LLE}_{\epsilon=0.05}$ stays finite as $N$ is increased and seems to eventually saturate to a finite value [see Fig.~\ref{LLEN:fig}(b)]. Here the dynamics shows chaos, but there is still synchronisation. In the chaotic-thermalisation regime, the behaviour of the LLE versus $N$ is very similar to the chaotic-synchronisation case and we do not show it.
\newpage

\begin{thebibliography}{100}
%
%\bibitem{Vulpiani}
%P.~Castiglione, M.~Falcioni, A.~Lesne, A.~Vulpiani,
%\newblock {\it Chaos and Coarse Graining in Statistical Mechanics} (Cambridge 2008).
%
\bibitem{Pikovsking}
A.~Pikovsky, M.~Rosenblum, and J.~Kurths,
\newblock {\it Synchronisation: A Universal Concept in Nonlinear Science} (Cambridge 2001).
%
\bibitem{Gupta}
S.~Gupta, A.~Campa, and S.~Ruffo,
\newblock {\it Statistical Physics of synchronisation} (Springer 2018).
%
\bibitem{arenas}
A.~Arenas, A.~D{\'i}az-Guilera , J.~Kurths, Y.~Moreno , and C.~Zhou
\newblock{ Physics Reports {\bf 469}, 93 (2008).}
%
\bibitem{Strogazz}
S. Strogatz,
\newblock {\it Sync: How Order Emerges from Chaos in the Universe, Nature, and Daily Life} (Penguin 2003).
%\bibitem{De_Luca}
%J.~De Luca, A.~J. Lichtenberg, and S.~Ruffo,
%\newblock {Phys. Rev. E {\bf 60}, 3781 (1999)}.
%
%\bibitem{Howard}
%J.~E. Howard, A.~J. Lichtenberg, and M.~A. Liebermann,
%\newblock{Physica {\bf 5D}, 243 (1982)}.
%
%
%\bibitem{Froesch}
%C.~Froeschl\'e, and J.-P. Sheidecker,
%\newblock{Phys. Rev. A, {\bf 12}, 2137 (1975)}.
%
%\bibitem{Miller}
%H.~L. Wright, and B.~N. Miller,
%\newblock{Phys. Rev. A, {\bf 29}, 1411 (1984)}.
\bibitem{Berry}
M.~V. Berry,
\newblock {\it Regular and Irregular Motion} in {\it Topics in Nonlinear Mechanics},
(ed. S. Jonna) {\bf 46}, 16-120, (1978).
%
\bibitem{Lichtenberg}
A.~J. Lichtenberg and M.~A. Lieberman,
\newblock{Regular and Chaotic Dynamics} ($2^{\rm nd}$ edition) (Springer-Verlag 1992).
%
\bibitem{Arnold_Avez}
V.~I. Arnold, and A.~Avez
\newblock {\it Ergodic Problems of Classical Mechanics} (Addison-Wesley 1989).
%
\bibitem{Pettini}
M.~Pettini, and M.~Landolfi,
\newblock{Phys. Rev. A {\bf 41}, 768 (1990)}.
%
\bibitem{Konishi}
T.~Konishi, and K.~Kaneko,
\newblock{J. Phys. A: Math. Gen. {\bf 23}, L715 (1990)}.
%
\bibitem{MBL_review}
D.~A. Abanin, E.~Altman, I.~Bloch, M.~Serbyn,
\newblock{Rev. Mod. Phys. {\bf 91}, 021001 (2019). }
%
\bibitem{Rozenbaum}
E.~B. Rozenbaum, and V.~Galitski,
\newblock{Phys. Rev. B {\bf 95}, 064303 (2017).}
%
\bibitem{Rylands}
C.~Rylands, E.~B. Rozenbaum, V.~Galitski, and R.~Konik,
\newblock{arXiv:1904.09473 (2019).}
%
\bibitem{Michele}
M.~Fava, R.~Fazio, and A.~Russomanno,
\newblock{arXiv:1908.03399 (2019).}
%
\bibitem{Tsuchiya}
T.~Tsuchiya, and N.~Gouda,
\newblock{Phys. Rev. E {\bf 61}, 948 (200)}.
%
\bibitem{Campa:book}
A.~Campa, T.~Dauxois, D.~Fanelli, and S.~Ruffo,
\newblock{\it Physics of Long-Range Interacting Systems} (Oxford 2014).
%
\bibitem{footnote}There may be also cases~\cite{Antoni}  where there are hints of a transition between regularity at 
high energy density~\cite{Latora,Firpo,Manos, Filho, Filho1} and chaoticity at low energy density. Generalisations 
of the Hamiltonian mean field model may display a crossover in the 
power-law exponent at large energy density from regularity to chaoticity~\cite{Celia,Firpo1}. 
%
\bibitem{Antoni}
M.~Antoni, and S.~Ruffo, 
\newblock{Phys. Rev. E {\bf 52}, 2361 (1995)}
%
\bibitem{Firpo}
M.-C. Firpo,
\newblock{Phys. Rev. E {\bf 57}, 6599 (1998)}.
%
\bibitem{Latora}
V.~Latora, A.~Rapisarda, and S.~Ruffo,
\newblock{Phys. Rev. Lett. {\bf 80}, 692 (1998).}
%
\bibitem{Manos}
T.~Manos, and S.~Ruffo,
\newblock{Transp. Theory Stat. Phys. {\bf 40}, 360 (2011)}.
%
\bibitem{Filho}
L.~H. Miranda Filho, M.~A. Amato, and T.~M. Rocha Filho,
\newblock{Jour. Stat. Mech. 033204 (2018)}.
%
\bibitem{Filho1}
T.~M. Rocha Filho, A.~E. Santana, M.~A. Amato, and A.~Figueiredo,
\newblock{Phys. Rev. E {\bf 90}, 032133 (2014)}.
%
\bibitem{Celia}
C.~Anteneodo, and C.~Tsallis,
\newblock{Phys. Rev. Lett. {\bf 80}, 5313 (1998)}.
%
\bibitem{Firpo1}
M.-C. Firpo, and S.~Ruffo,
\newblock{J. Phys. A: Math. Gen. {\bf 34}, L511 (2001)}.
%
\bibitem{Oven}
%
O.~Howell, P.~Weinberg, D.~Sels, A.~Polkovnikov, and Marin Bukov,
\newblock {Phys. Rev. Lett. {\bf 122}, 010602 (2019).}.
%
%\bibitem{Nayak2}
%F.~Machado, G.~D. Meyer, D.~V. Else, C.~Nayak, N.~Y. Yao,
%\newblock {arXiv:1708.01620 (2017)}.
%
\bibitem{Else}
D.~V. Else, B.~Bauer, and C.~Nayak,
\newblock {Phys. Rev. Lett. {\bf 117}, 090402 (2016)}.
%
\bibitem{Vedika}
V.~Khemani, A.~Lazarides, R.~Moessner, and S.~L. Sondhi,
\newblock {Phys. Rev. Lett. {\bf 116}, 250401 (2016)}.
%
\bibitem{Chetan}
N.~Y. Yao, C.~Nayak, L.~Balents, and M.~P. Zaletel,
\newblock {arXiv:1801.02628 (2018).}
%
\bibitem{Gambetta}
F.~M. Gambetta, F.~Carollo, A.~Lazarides, I.~Lesanovsky, and J.~P. Garrahan,
\newblock {arXiv:1905.08826 (2019).}

%
\bibitem{fabri}
The concept of broken-symmetry edge was introduced in: G.~Mazza, and M.~Fabrizio,
\newblock{Phys. Rev. B {\bf 86}, 184303 (2012).}
%
\bibitem{FTC_Russomanno}
A. Russomanno, F. Iemini, M. Dalmonte, and R. Fazio,
\newblock {Phys. Rev. B {\bf 95}, 214307 (2017)}.
%
%\bibitem{federica}
%F.~M. Surace, A.~Russomanno, M.~Dalmonte, A.~Silva, R.~Fazio, and F.~Iemini,
%\newblock {Phys. Rev. B {\bf 99}, 104303 (2019).}
%
\bibitem{kuramoto}
Y.~Kuramoto in {\it International Symposium on Mathematical Problems in Theoretical Physics} (ed. H.~Araki), {\it Lecture Notes in Physics}, {\bf 39}, 420, (Springer-Verlag 1992).
%
\bibitem{kuramoto1}
Y.~Kuramoto
\newblock {\it Chemical Oscillations, Waves, and Turbulence} (Springer-Verlag 1984).
%
%\bibitem{cross}
%D.~Chowdhury, and M.~C. Cross,
%\newblock{Phys. Rev. E {\bf 82}, 016205 (2010).}
%
\bibitem{Ekkekkac}
M. Kac, J. Math. Phys. {\bf 4}, 216 (1963).
%
\bibitem{chiri_vov}
B.~V. Chirikov and V.~V. Vecheslavov,
\newblock{Jour. Stat. Phys. {\bf 71}, 243 (1993).}
%
\textcolor{black}{
\bibitem{Ata_ema}
A.~Rajak, I.~Dana, and E.~G. Dalla Torre,
\newblock{Phys. Rev. B {\bf 100}, 100302(R) (2019).}
}
%
\bibitem{Haake}
F.~Haake, M.~Ku\'s, and R.~Scharf,
\newblock{Z. Phys. B {\bf 65}, 381 (1987).}
%
\textcolor{black}{
\bibitem{mori}
T.~Mori,
\newblock{J. Phys. A: Math. Theor. {\bf 52}, 054001 (2019).}
%
\bibitem{gorshkov}
F. Liu, R. Lundgren, P. Titum, G. Pagano, J. Zhang, C. Monroe, and A.~V. Gorshkov,
\newblock{Phys. Rev. Lett. {\bf 122}, 150601 (2019).}
}
%
%\bibitem{Ott}
%E.~Ott,
%\newblock {\it Chaos in Dynamical Systems} (Cambridge 2002).
%
\bibitem{Kolmogorov_entropy}
G. Benettin, L. Galgani, and J. M. Strelcyn,
\newblock {Phys. Rev. A. {\bf 14}, 2338 (1976)}.
%
\bibitem{Lyapunov_book}
A.~Pikovsky and A.~Politi,
\newblock {\it Lyapunov exponents: a tool to explore complex dynamics } (Cambridge 2016).
%
\bibitem{PRE}
O.~V. Popovych, Yu.~L. Maistrenko, and P.~A. Tass,
\newblock Phys. Rev. E {\bf 71}, 065201(R) (2005).
%
\textcolor{black}{
\bibitem{boris}
A.~Patra, B.~L. Altshuler and E.~A. Yuzbashyan,
\newblock{Phys. Rev. A {\bf 100}, 023418 (2019)}
}
%
\bibitem{lukin}
W.~W. Ho, S.~Choi, M.~D. Lukin, and D.~A. Abanin,
\newblock {Phys. Rev. Lett. {\bf 119}, 010602 (2017)}.
%
\bibitem{mukamel2}
C.~H. Bennett, G.~Grinstein, Y.~He, C.~Jayaprakash, and D.~Mukamel,
\newblock Phys. Rev. A {\bf 41}, 1932 (1990).
%
\bibitem{mukamel1}
G.~Grinstein, D.~Mukamel, R.~Seidin, and C.~H. Bennett,
\newblock Phys. Rev. Lett. {\bf 70}, 3607 (1993).
%
\end{thebibliography}
\end{document}